\RequirePackage{fix-cm}
\documentclass[smallcondensed]{svjour3}     % onecolumn (ditto)
\smartqed  % flush right qed marks, e.g. at end of proof
\usepackage{graphicx}
\usepackage{multirow}

\journalname{}

\begin{document}

\title{The Boyer-Moore Waterfall Model Revisited}

\author{Petros Papapanagiotou         \and
        Jacques Fleuriot %etc.
}
 
\institute{Petros Papapanagiotou \at
              Center of Intelligent Systems and their Applications\\
					    School of Informatics, The University of Edinburgh\\
					    Informatics Forum, 10 Crichton Street, Edinburgh EH8 9AB, UK.  \\
              Tel.: +44 (0)131 651 3077\\
              \email{pe.p@ed.ac.uk}           %  \\
           \and
           Jacques Fleuriot \at
              Center of Intelligent Systems and their Applications\\
					    School of Informatics, The University of Edinburgh\\
					    Informatics Forum, 10 Crichton Street, Edinburgh EH8 9AB, UK.  \\
              \email{jdf@inf.ed.ac.uk}
}

\date{08 September 2011}

\maketitle

\begin{abstract}
In this paper, we investigate the potential of the Boyer-Moore waterfall model for the automation of inductive proofs within a modern proof assistant. We analyze the basic concepts and methodology underlying this 30-year-old model and implement a new, fully integrated tool in the theorem prover HOL Light that can be invoked as a tactic. We also describe several extensions and enhancements to the model. These include the integration of existing HOL Light proof procedures and the addition of state-of-the-art generalization techniques into the waterfall. Various features, such as proof feedback and heuristics dealing with non-termination, that are needed to make this automated tool useful within our interactive setting are also discussed.  Finally, we present a thorough evaluation of the approach using a set of 150 theorems, and discuss the effectiveness of our additions and relevance of the model in light of our results.

\keywords{interactive theorem proving \and induction \and Boyer-Moore \and waterfall model \and HOL Light}
\end{abstract}

\section{Introduction}
\label{Intro}

Boyer and Moore's seminal book ``A Computational Logic'' \cite{boyer1979cl} covered in detail the most important aspects of the design of an automated theorem prover based on a ``waterfall'' model. In particular, it focused on recursive data types and functions, and, consequently, on proofs by induction. 
A lot of the ideas from this detailed work are still being used in modern research for automated inductive proofs. For example, the Nqthm system \cite{nqthm}, which started off as an implementation of a similar model to Boyer-Moore's original prover, later evolved into ACL2, system which is still under development \cite{kaufmann2000car}. Although ACL2 is now a much more sophisticated and powerful system than the original Boyer-Moore waterfall approach, we wanted to investigate whether this venerable model could still be beneficial to  modern, general-purpose theorem proving systems.

Our investigation involves the integration of the Boyer and Moore waterfall model into the HOL Light theorem prover, followed by its extension with modern algorithms and procedures. Our work reconstructs Boulton's implementation of the Boyer-Moore system \cite{boulton1992bma} from HOL 90 (an earlier version of HOL), which is believed to be a quite faithful reconstruction of the Boyer-Moore approach.

The paper is organized as follows: In Section \ref{HOLLight}, we briefly discuss HOL Light, a state-of-the-art theorem prover. In Section \ref{sec:BM}, we review the waterfall model as it was originally suggested by Boyer and Moore. This is followed, in Section \ref{sec:Port}, by the details of our implementation and the extensions that we added, including state-of-the-art generalization algorithms. In Section \ref{sec:Eval}, we analyze the setup and results of the system evaluation. A brief review of related work in included in Section \ref{sec:Related}. We describe our suggestions for future work in Section \ref{sec:Future} and summarize our conclusions in Section \ref{sec:Concl}.

\section{HOL Light}
\label{HOLLight}
HOL Light \cite{harrison1996hdr} is a relatively recent member of the HOL family of theorem provers that was initially built in an attempt to overcome certain disadvantages of its predecessors. 

The system has equality as the only primitive concept and a few primitive inference rules that form the basis of more complex rules and tactics. Built on top of these, HOL Light has its own automated methods for proofs such as the model elimination method MESON \cite{harrison1996ops}. Additionally, it has an array of \emph{conversion} methods that allow for efficient and fine-grained manipulation (such as rewriting or numerical reduction) of formulas. 

HOL Light has significant advantages over the other modern systems especially for the current work. It is a lightweight, flexible system written in OCaml, that allows for interaction at every level. This allows for relatively easy implementation and integration of tools that can seamlessly interact with the internals and methods of HOL Light. 

Unfortunately, there are also a few disadvantages. HOL Light is not too user-friendly when writing proofs due to its relatively large number of low-level tactics and complicated syntax. It has a steep learning curve and its procedural proofs have reduced readability compared to systems such as Isabelle \cite{paulson1994igt}, where the declarative proof-style seems now to be the norm. In a nutshell, HOL Light can be characterised as a system functioning at a lower \emph{programming} level rather than the higher but limiting user level. For our purpose though, the advantage of a smooth and direct interaction, coupled with the fact that HOL Light is a well-regarded and powerful system were convincing enough to select it as the backend for the Boyer-Moore system.

\section{The Boyer \& Moore Model}
\label{sec:BM}

In the next few sections, we provide a review of the Boyer-Moore waterfall model. In particular, we describe its architecture and the various heuristics that are applied in the automatic search for a proof. We note that that this description encompasses both the original model and Boulton's HOL reconstruction, which, as mentioned before, is believed to be mostly faithful. Nevertheless, we shall point out any aspects, where Boulton's HOL version seems to diverge slightly from the original model either by design or due to the use of the HOL system as a vehicle.

\subsection{The Waterfall Metaphor}
\label{sec:Waterfall}
One of the main principles underlying the Boyer-Moore model is the application of ``black-box'' procedures. According to Boyer and Moore, induction should be applied only as a last resort when all the other procedures have failed. Moreover, one must ensure that induction is applied to the simplest and most general clauses. The black-box procedures either prove or, failing that, simplify and generalize the clauses as much as possible so as to prepare them for induction. It should be noted that Boyer and Moore call all such procedures ``heuristics'' even though not all of them use heuristic methods and, for this paper, we shall follow their terminology.

\begin{figure}[htbp]
	\centering
		\includegraphics{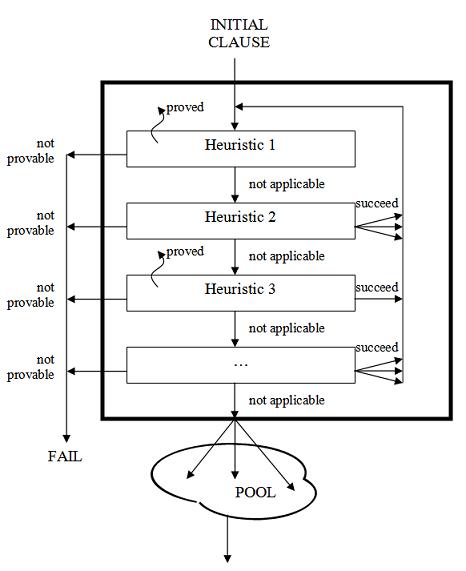}
	\caption{Diagram of the Waterfall model}
	\label{fig:waterfall}
\end{figure}

The heuristics are organised and applied in a way that metaphorically resembles rocks in an initially dry waterfall. Clauses that are to be proven are poured from the top of the waterfall and each heuristic is then applied to the clause sequentially. The application of each heuristic can have one of the following results:
\begin{itemize}
	\item It may prove a clause, in which case the latter `evaporates'.
	\item Sometimes it may simplify or split the clause into smaller ones. In this case, the proof of the resulting clauses is sufficient for the proof of the initial clause. We then say that the heuristic has been \emph{applied successfully} or simply was \emph{successful}. The newly created clause or clauses are recursively poured from the top of the waterfall. 
	\item It may disprove the clause, for example by reducing it to False, in which case the system immediately fails.
	\item If it cannot deal with the clause, it passes it on to the next heuristic. In that case we say that the heuristic has \emph{failed}.
\end{itemize}
    If all the heuristics fail, the clause ends up at the bottom of the waterfall and, together with all the clauses that the waterfall failed to prove, forms a pool. The aim then is to prove each clause of the pool by induction. Doing so is sufficient to prove the initial conjecture. An illustration of the model we have just described can be found in Fig. \ref{fig:waterfall}.

\begin{figure}[htbp]
	\centering
		\includegraphics{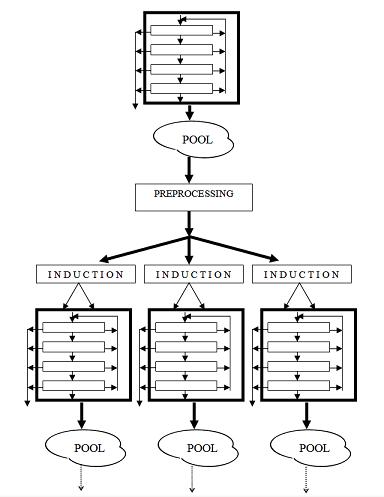}
	\caption{The Boyer-Moore Model}
	\label{fig:waterdia}
\end{figure}

Once induction is applied to one of the clauses in the pool, the newly produced clauses (the base case and step case) are in turn poured over a \emph{new} waterfall. The same process of heuristic application as before is then used. New pools of clauses may be formed and another induction may be applied to them as illustrated in Fig. \ref{fig:waterdia}. Eventually, assuming the system is successful, all clauses will be proved and will have evaporated from all pools and waterfalls, resulting in the proof of the initial clause.

\subsection{The Shell}
\label{sec:Shell}

The ``Shell principle'' \cite{boyer1979cl} was used in the original Boyer-Moore model to define and describe recursive datatypes. Such a principle was crucial to this initial description because of the lack of support for such datatypes in Lisp, the underlying programming language for the system. Boulton, in his HOL90 implementation, uses an extended version of the original Shell that contains more defined properties and information. Even though the HOL systems, including HOL Light, have full support for recursive data types, the implementation and usage of the Shell within the automated system is still necessary. This is because it contains useful, explicitly defined information about the types that needs to be readily available to the automated waterfall heuristics at any given point.

Boyer and Moore describe a Shell as a ``colored $n$-tuple with restrictions on the colors of objects that can occupy its components'' \cite{boyer1979cl}. The color represents a unique identifier for a datatype. Apart from identifying the datatype and separating it from other similar data types, a number of properties are defined for it within the Shell. For example it will contain constructors, bottom objects and accessors (also known as ``destructors'' in the more recent literature) as some of its main parts. Boulton, in his HOL versions, included additional properties such as a type axiom, an induction theorem, a theorem for splitting cases, and theorems to ensure distinctness of constructors and one-one restrictions. As an example, we provide the shell for natural numbers (type \emph{num} in HOL Light) in Table \ref{tab:ShellEx}.

\begin{table}
\caption{Shell for Natural Numbers}
\label{tab:ShellEx}      
\begin{tabular}{rl}
\hline\noalign{\smallskip}
Name: & ``$num$''\\
Arguments: & []\\
\noalign{\smallskip}\hline\noalign{\smallskip}
Bottom Object: & $0$\\
Constructors: & $SUC$ $(num)$ \\
\noalign{\smallskip}\hline\noalign{\smallskip}
Accessors: & $PRE$ : $\vdash\ \forall n.\ PRE\ (SUC\ n)\ =\ n$ \\
Type Axiom: & $\vdash\ \forall e\ f.\ \exists g.\ g\ 0\ =\ e\ \wedge\ (\forall n.\ g\ (SUC\ n)\ =\ f\ (g\ n)\ n)$ \\
Induction theorem: & $\vdash\ \forall P.\ P\ 0\ \wedge\ (\forall n.\ P\ n\ \Rightarrow\ P\ (SUC\ n))\ \Rightarrow\ (\forall n.\ P\ n)$ \\
Cases theorem: & $\vdash\ \forall m.\ m\ =\ 0\ \vee\ (\exists n.\ m\ =\ SUC\ n)$ \\
Distinctness theorem(s): & $\vdash\ \forall n.\ \neg(SUC\ n\ =\ 0)$ \\
One-one restriction(s): & $\vdash\ \forall m\ n.\ SUC\ m\ =\ SUC\ n\ \Leftrightarrow\ m\ =\ n$ \\
\noalign{\smallskip}\hline
\end{tabular}
\end{table}

\subsection{The heuristics}
\label{sec:Heuristics}
There are seven heuristics proposed in the original Boyer-Moore system. These include the transformation to clausal form and the induction heuristic which applies the induction scheme. As mentioned previously, the induction heuristic is applied separately from the waterfall loop, but it is still implemented using the same structure and output as all the other heuristics. We will describe the six heuristics that form part of the waterfall next, focusing on their functionality, limitations and output.

\subsubsection{The Clausal Form Heuristic}
\label{sec:CLF}
Boyer and Moore decided to rely on Clausal Normal Form (CNF) because they could avoid an asymmetry they observed with conditionals. Generally, a term can be transformed to CNF (i.e. a conjunction of disjunctions) by eliminating existential quantifiers through Skolemisation and removing universal quantifiers. Each conjunct is then a disjunction of literals and is called a clause. For example, the term $m\;+\;n\;=\;0\ \Rightarrow\ m\;=\;0\;\wedge\;n\;=\;0$ is transformed into $(\neg(m\;+\;n\;=\;0)\;\vee\;m\;=\;0\;)\; \wedge \;(\neg(m\;+\;n\;=\;0)\;\vee\;n\;=\;0\;)$ which is a conjunction of two clauses. 

In the Boyer-Moore model, the Clausal Form heuristic is responsible for the transformation of \emph{quantifier-free} sentences to CNF. It fails if the input term is a single clause already in CNF. It also splits a conjunction of clauses and returns them as a list. We note that an important limitation of this heuristic as implemented in the Boyer-Moore system is that it cannot deal with quantifiers, ie. it assumes quantifiers have already been eliminated.

\subsubsection{The Substitution Heuristic}
\label{sec:Subst}
The Substitution heuristic is a simplification procedure used to eliminate negations of equalities between variables and terms. For example, assume we have the following input, where $x$ is a variable and $A_{1}$, $A_{2}$, $A_{3}$ do not contain $x$:
	\[A_{1}\ \vee\ \neg(x\ =\ t)\ \vee\ A_{2}\ \vee\ P(x)\ \vee\ A_{3}\]
If $t$ is a term that does not contain $x$ as a variable then we can substitute $x$ in $P(x)$ with $t$, thus obtaining:
	\[A_{1}\ \vee\ F \vee\ A_{2}\ \vee\ P(t)\ \vee\ A_{3}\]
We note that the negations of equalities often appear in CNF because such equalities are often on the left-hide side of an implication, either as part of the initial conjecture as a by-product of induction.

The heuristic fails if it cannot be applied, meaning that there is no such negated equality. Otherwise, the heuristic returns a single simplified clause.

\subsubsection{The Simplify Heuristic}
\label{sec:Simp}
This heuristic applies rewriting to the clause in an attempt to simplify or prove it. It uses rewrite rules defined by the user (see Section \ref{sec:UI}), definitions of recursive functions, and a few special rules for specific cases of clauses, such as conditionals (eg. the rule $if\ p\ then\ q\ else\ q\  \Leftrightarrow\ q$). The heuristic fails if no rules can be applied, or if no changes are made to the clause. It also uses lexicographic ordering in an attempt to avoid looping that may be caused by permutative rules and supports conditional rewrite rules. Unfortunately, this does not eliminate all possible loops and it is left to the user to ensure the set of rewrite rules is terminating. The methods to manipulate the set of rules are rather limited: they only allow the creation of new rules from existing, proved theorems and there is no explicit mechanism to remove a rule from the set.

\subsubsection{The Equality Heuristic}
\label{sec:Eq}
The Equality heuristic is similar to the substitution one. It uses equalities for ``cross'' (or ``weak'') fertilization. Cross-fertilization is the replacement of part of the induction hypothesis within the induction conclusion (as opposed to a complete replacement in strong ``fertilization''). Instead of negations of equalities between a variable and terms not containing the variable as in the substitution heuristic, the equality heuristic checks for negations of equalities involving a term which is not a so-called  \emph{explicit value template}. An explicit value template is a non-variable term consisting of constants or any constructor application to bottom objects or variables. For example $0$, $SUC(0)$ and $SUC(SUC(x))$ are all explicit value templates. If the clause is the result of an induction step, the negated equality is eliminated (because of cross-fertilization). The heuristic fails if no such negated equality is found. As an example, during the proof of the commutative property of multiplication we obtain the following clause as an induction step:  $\neg(n\times 0\;=\;0)\ \vee\ (n\times 0)\;+\;0\;=\;0$. The equality heuristic is applied in this case, as $n\times 0$ is not an explicit value template, giving us the result: $F\ \vee\ 0+0\;=\;0$ ie. $0+0\;=\;0$.

\subsubsection{The Generalization Heuristic}
\label{sec:Gen}
The Generalization heuristic attempts to substitute the clause with a more general one that might be easier to prove. In particular, the generalization proposed by Boyer and Moore, and thus implemented by Boulton, is based on the elimination of minimal common subterms.

First, the generalizable terms of the clause are calculated. A term is generalizable if it is neither a variable, nor an explicit value template (see section \ref{sec:Eq}), nor an application of accessor functions. From the generalizable terms, candidates for generalization are picked based on the common subterm criterion. According to this criterion, a generalizable term is a candidate for generalization if it appears in more than one generalizable subterms, or on both sides of an equation, or on both sides of a negated equation. Finally, from the list of candidates, the minimal common subterms are picked, meaning that candidates that have other candidates as subterms are rejected. Thus the ``smallest'' candidates are generalized simultaneously. These subterms are replaced with fresh variables. The heuristic fails if no such subterms are found. Otherwise, it returns a single generalized clause. The original clause follows by a simple instantiation of the variable in the generalized clause.

As an example, taken from an actual test case, when trying to prove the commutative property for the multiplication of natural numbers, the system ends up with the clause $(m\ \times\ n)\ +\ n\ =\ n\ +\ (m\ \times\ n)$ after a few steps. In this clause, the term $(m\ \times\ n)$ is on both sides of the equation and may be generalized and substituted with a new variable $n'$. The resulting clause is $n'\ +\ n\ =\ n\ +\ n'$.

It is worth noting that generalization produces new clauses, some of which can be particularly interesting. In our example, generalization yields the commutative property of addition for natural numbers. As a result, generalization is considered to be a form of lemma speculation.

Unfortunately, as expected, the process is not flawless. It may over-generalize, resulting in a clause which is no longer provable. In other words, it is sometimes the case that the original clause might have been provable if we just had proceeded with induction rather than generalization. Creating a generalization process that minimizes the risk of over-generalizing remains an open issue.

The heuristic additionally supports the use of generalization lemmas supplied by the user. These are theorems that ``point out properties of terms that are good to keep in mind when generalizing formulas'' \cite{boyer1979cl}. If one of the candidate subterms is an instantiation of the generalization lemma, the lemma is added to the clause so as to ``keep in mind'' the property that it represents. For instance, if we use the (trivial) generalization lemma $m\ \times\ n\ \geq\ 0$ then, in our last example, after generalization, we obtain the additional restriction $n'\geq0$ and the result is $n'\geq0\ \Rightarrow \ n'\ +\ n\ =\ n\ +\ n'$. We should note that this result is not in CNF but will be converted in the next proof step, as it will be poured at the top of the waterfall and go through the Clausal Form Heuristic (see Section \ref{sec:CLF}).

\subsubsection{The Irrelevance Heuristic}
\label{sec:Irr}
This heuristic is another form of generalization. It attempts to eliminate irrelevant subterms from the clause. Firstly the subterms of the clause are split into partitions based on common variables, meaning that two subterms are in the same partition if they share at least one variable. One such partition is irrelevant if it is falsifiable. Judging if a partition of subterms is falsifiable is done using two actual heuristics. The first one checks if there are any occurrences of recursive functions. If not, then the subterms consist only of functions of the shell (constructors, accessors, constants etc). Therefore, if the partition was always true, we should have proved it by simplification. Since the irrelevance check comes after the simplifier in the waterfall, we have certainly failed to do so and consequently we can assume that the partition of subterms can be falsified. The second heuristic checks if a subterm is an application of a function over variables. If so, it can only be a theorem if the function always returns true. Again it is assumed that it should have been simplified by rewriting. If a partition of subterms can be falsified, it is safe to eliminate the subterms from the clause. The resulting clause will be a theorem if and only if the original one is a theorem. 

We illustrate the above idea using an example. Consider the clause:
\[\mbox{\fontsize{6}{8}\selectfont $p\ =\ []\ \vee\
REVERSE\ (APPEND\ (REVERSE\ p)\ [a])\ =\ CONS\ a\ (REVERSE\ (REVERSE\ p))$}\]
\normalsize which is generalized to: 
\[\mbox{\fontsize{6}{8}\selectfont $p\ =\ []\ \vee
REVERSE\ (APPEND\ l\ [a])\ =\ CONS\ a\ (REVERSE\ l)$}\] 
\normalsize In this result, the subterm $p\ =\ []$ is deemed irrelevant because it does not have common variables with the rest of the term, does not have any application of recursive functions nor is an application of a function to variables. After eliminating the irrelevant term, the resulting clause is: 
\[\mbox{\fontsize{6}{8}\selectfont $REVERSE\ (APPEND\ l\ [a])\ =\ CONS\ a\ (REVERSE\ l)$}\] \normalsize which is a generalization of the original one.

Unfortunately, these heuristics are unsafe and may eliminate relevant subterms, thereby rendering the clause unprovable. The heuristic fails if no irrelevant terms are found, or it indicates that the clause cannot be proved if it finds that all subterms are irrelevant. Otherwise, it returns a simplified clause and the proof of the original one.

\subsection{User Interaction}
\label{sec:UI}
However systematic the system that we describe might be, it still does not guarantee to find the proof of a true statement, i.e. it is not complete. Thus, it may require some user intervention to ``set the tracks'' and guide the proof procedures. The user can interact with the system and affect its performance in various simple ways:
\begin{itemize}
	\item Firstly, the user is responsible for providing the shell for the data type and the definitions of the functions, both simple and recursive.
	\item Moreover, the user can manipulate the sets of rewrite rules and generalization lemmas. Picking the set of rewrite rules carefully may prove crucial for achieving the proof. Allowing the user to manipulate the set offers significant control over the proof procedure.
	\item Additionally, picking generalization lemmas (see Section \ref{sec:Gen}) containing useful properties may help guide or unlock proofs that would otherwise fail.
	\item The user may also choose which main waterfall heuristics will be used and in what order. Different combinations of heuristics may produce different results. For instance, the user may choose to remove the generalization heuristic which, as an unsafe operation, might over-generalize and render a conjecture unprovable. 
\end{itemize}

\section{The Boyer-Moore Waterfall Model implemented and extended in HOL Light}
\label{sec:Port}
In this section we discuss our implementation of the Boyer-Moore model in HOL Light. The main system consists of a reimplementation of Richard Boulton's old code HOL90 \cite{boulton1992bma}. The main issues of this reimplementation are discussed in Section \ref{sec:Issues}. We also proceeded to develop various enhancements and improvements to the system in our attempt to evaluate its potential and effectiveness within a state-of-the-art theorem prover. The enhancements were applied in number of steps. Firstly, we made some effort to fix some issues and upgrade the system so that it is a better fit to our current interactive setting. These changes are discussed in Section \ref{sec:Fitting}. Secondly, we focused on integrating some of HOL Light's features into the system and these attempts are analysed in Section \ref{sec:Exploit}. Finally, as will be described in Section \ref{sec:Aderhold}, we attempted to upgrade the generalization heuristics by introducing some of the latest work in this area. Moreover, in an attempt to address over-generalization issues we implemented and integrated a simple disprover, which is discussed in Section \ref{sec:Counter}

\subsection{Main issues}
\label{sec:Issues} 
The primary implementation task was to reconstruct the old code by \cite{boulton1992bma} for HOL Light. It is important to note that this was not a simple, straightforward translation from one environment to another, since HOL Light has significant differences from HOL90, and the systems lack complete documentation. The encountered issues can be split into two basic categories, which we briefly discuss.

i) The first one involves those caused by the differences between Standard ML (SML) used to implement HOL90 and OCaml used of HOL Light. There are syntactic variations, such as those in function and data type declarations, in case splits, in the test for the empty list (equality to an empty list is used instead of the $null$ function) and many more.  Combined with the limited documentation for these platforms (consisting mainly of expert users offering solutions through mailing lists), these made some aspects of the re-implementation task a tedious process. Dealing with logical differences and resulting errors was even harder.

ii) The second category involves those caused by the difference in system functionalities. For example, some inference rules and tactics that existed in HOL90 have no counterparts in HOL Light. For instance, ``SUBS\_OCCS'' (a rule used to substitute occurrences of a term in a theorem using other equational theorems) and ``INDUCT\_TAC'' (a tactic used to apply induction based on a given induction rule). We were compelled to reconstruct the missing rules and tactics based on the existing ones in HOL Light. Differences in the system behaviour also had an impact on the reconstruction. As an example, HOL Light treats natural numbers not as constants (as is the case in HOL) but as applications of the $NUMERAL$ function.

\subsection{Fitting the model into an interactive setting}
\label{sec:Fitting}
The first challenge that we encountered, once the initial reconstruction of the system had been accomplished, involved augmenting the means of user interaction so as to improve the fit of the automatic system within HOL Light's interactive setting. The two main steps we took towards this goal were the extension and improvement of the feedback provided by the system (Section \ref{sec:Verbosity}) and the attempt to minimize non-termination (Section \ref{sec:Loops}).

\subsubsection{Increasing the System Verbosity}
\label{sec:Verbosity}
One of the simplest, yet important, issues involved in testing, evaluating and improving the system is to have the option of producing a trace for every proof attempt. In order to investigate the reasons for failure or error, one naturally desires as much information as possible. However, this need has to be balanced against the fact that too much information may lead to clutter when dealing with large proofs, making the trace unreadable.

Boulton's original implementation of the system contained a minimalistic proof printer. Upon activation, it would give information about which clause is being evaluated by the waterfall at any given time. We enhanced this proof printer so as to offer richer information about the mechanics of the system. Each heuristic, upon success, prints out its name before the resulting clause. Many offer even more information about their results. For example the Clausal Form heuristic shows the number of new clauses produced (by breaking conjunctions) and the Generalization heuristic shows which subterms were generalized. A message is also printed out whenever induction is applied, indicating the clause to which it was applied. Therefore, the steps followed in the proof process are now made explicit. Moreover, a machinery was included to indicate the reason for failure, wherever possible, as well as the theorem produced upon the successful proof of a clause. Finally, the user was given the option of viewing the proof tree created by the waterfall upon its completion and before moving to induction. As the proof trace may increase drastically when dealing with complicated theorems, we aimed to keep the messages compact and easy to read. The improved tracing mechanism effectively gave us means of properly monitoring the system, when need be, and of analyzing its performance and finding solutions to its problems. As a simple example, the proof trace for the simple lemma $SUC(m)\ =\ m\ +\ SUC(0)$ is given in Fig. \ref{fig:trace}.
\begin{figure}[htbp]
		\ttfamily
		\begin{verbatim}
SUC m = m + SUC 0
Doing induction on:SUC m = m + SUC 0


 SUC 0 = 0 + SUC 0
-> HL Simplify Heuristic
Proven:|- SUC 0 = 0 + SUC 0


 SUC n = n + SUC 0 ==> SUC (SUC n) = SUC n + SUC 0
-> Clausal Form Heuristic (1 clause)
 ~(SUC n = n + SUC 0) \/ SUC (SUC n) = SUC n + SUC 0
-> HL Simplify Heuristic
 ~(SUC n = n + SUC 0) \/ SUC n = n + SUC 0
-> Tautology Heuristic
Proven:|- ~(SUC n = n + SUC 0) \/ SUC n = n + SUC 0


val it : thm = |- SUC m = m + SUC 0
		\end{verbatim}
	\caption{A Sample Proof Trace}
	\label{fig:trace}
\end{figure}

\subsubsection{Eliminating Loops}
\label{sec:Loops}
One of the first disadvantages of the original Boulton implementation, as noticed during its reconstruction in HOL Light, was that the system would in some cases fall into endless loops. This is a tricky issue for such automated systems because the user is in no position of knowing if progress is being made towards the proof or if the system will never terminate. This becomes particularly troublesome when the system is used as part of an extensive run on a set of hundreds of theorems. As a way of dealing with this issue, we decided to introduce two techniques: a warehouse filter and the imposition of a maximum depth limit on the size of terms. These are applied outside the waterfall model, as described next.

The warehouse filter is a storage of clauses that have already been evaluated successfully by a given waterfall. If the same clause is poured on top of the same waterfall it means that at least one of the heuristics was successful but after one or more loops the system ended up with the same result. Consequently, if we allow it to proceed further, the same heuristic will be applied and the same result will loop through the waterfall forever. Our filter checks if the clause has already been evaluated by the waterfall and which heuristic was applied to it. It then skips the heuristic that lead to the loop and tries the next one instead in the hope of eventually achieving the proof. It is worth noting that the warehouse is local. Therefore, if the same clause is poured over a different waterfall (eg. after at least one induction step) it will not be filtered, as it is not certain in that case that there is a loop. For example it might just be a subterm that occurs more than once in the same proof. The same warehouse filtering technique is also applied in the induction scheme. Before applying induction we check if induction has already been applied to the same clause in the same proof branch. If this is the case, the system fails because further induction will only lead to the same result.

Despite our efforts with the warehouse filter and its effectiveness in some situations, it was still insufficient as the system still looped fairly often. After careful observation of various non-terminating cases, we noticed that in most of them, the repetitive application of rewriting and inductions lead to a constant increase of the size of the term by having multiple constructors or function applications to a variable. Our ``maximum depth'' heuristic measures the maximum depth in the syntax tree of a term where a variable occurs. By adding a user-defined limit to this depth we accomplished a drastic decrease in the number of looping cases (see Section \ref{sec:Eval} for detailed results and evaluation). Unfortunately, it is possible for the heuristic to interrupt proofs that might eventually succeed. However, given our interactive environment, early termination was favoured over lengthy proof times. Moreover, despite this heuristic, not all loops were eliminated. In some cases, for example, the terms can expand very slowly (more than 10 minutes to reach maximum depth limit in some of our evaluation tests). In other cases, one of the terms kept being split into multiple clauses after being rewritten. Investigating more heuristics to tackle these cases or more sophisticated techniques used in similar automated systems (such as incremental depth search used in HOL Light's MESON tactic) is part of future work (see Section \ref{sec:Future}).

\subsection{Integrating HOL Light tools}
\label{sec:Exploit}
In this section, we discuss the integration of two HOL Light tools into the waterfall and some of the resulting issues. In particular, we tried to exploit HOL Light's tautology prover and simplifier within our system.

\subsubsection{The Tautology heuristic}
\label{sec:Taut}
HOL Light includes an automated procedure that can be used to prove tautologies. It can successfully deal with terms such as $p \vee \neg p$, $p=p$ and $(p \Rightarrow q) \vee (q \Rightarrow p)$ where $p$ and $q$ are atomic formulas that are not necessarily propositional. We exploited this function to build a tautology heuristic for the waterfall. The heuristic is placed at the very top of the waterfall for maximum efficiency since it does not alter the clause in any way, it only proves the clause immediately if it can.

\subsubsection{The HOL Light Simplifier}
\label{sec:HLsimp}
HOL Light's simplifier is a powerful and efficient tool, which is the workhorse for many proofs. In an attempt to exploit the efficiency of this simplifier in our system, we we devised a version of the system in which we replaced the simplify heuristic with one of HOL Light's conversions, the so-called REWRITE\_CONV. The new simplify heuristic works in a similar way to the original one (see Section \ref{sec:Simp}). One of the major differences, though, is that the original simplifier only rewrote recursive functions based on their definitions. The new heuristic is allowed to apply all rules (ie. both derived rewrite rules and definitions) at all times. Such a behaviour, we did realise, could be both an advantage, as it might provide more powerful simplification in some cases, and a disadvantage, because of the increased likelihood of looping.

\subsubsection{The Setify heuristic}
\label{sec:Setify}
The use of HOL Light's simplifier required a new, straightforward heuristic to deal with an issue that the original Boyer-Moore simplifier dealt with as one of its steps. In some cases, after several proof steps, a clause may end up including the same subterm as a disjunct more than once. The original Boyer-Moore simplifier would then remove such duplications, keeping only one copy of any disjunct in a clause. To achieve the same behaviour, we therefore created a heuristic to simplify such clauses by eliminating duplicate disjuncts. So, for a clause such as $A \vee B \vee A$, the second $A$ term is eliminated giving $A \vee B$ as a result. The heuristic also helped prevent some loops where a clause would endlessly expand with multiple identical disjuncts.

Next, our attempts focused on the improvement of the original Boyer-Moore generalization heuristics using some state-of-the-art techniques described in Section \ref{sec:Aderhold}. Finally, we made an attempt on a simple counterexample checker, described in Section \ref{sec:Counter}, which allowed us to avoid several overgeneralizations.

\subsection{Incorporating state-of-the-art Generalization techniques}
\label{sec:Aderhold}

Recent research on formula generalization has provided better heuristics and more filters to avoid over-generalizations. In particular, we studied Aderhold's approach, which is summarized in a recent paper \cite{aderhold}. In this work, a generalization heuristic and a tactic are created for a verification system called VeriFun \cite{walther2002vfb} and are shown to be effective at dealing with a substantial range and number of inductive properties. Aderhold's research builds on well-regarded generalization mechanisms, such as the ones used in the Boyer-Moore system, as well as novel ideas.

Aderhold's generalization heuristic contains five subprocesses, each handling a different aspect of generalization and not all of which are applicable to our system. We chose to implement the generalization of common subterms as an alternative for the generalization heuristic in the waterfall. We also implemented the algorithm for generalizing variables apart. 

We note that for any comparison between Alderhold's and the original waterfall algorithms within the Boyer-Moore model, that one should bear in mind that Aderhold's techniques are applied in a system with \emph{destructive-style} rather than \emph{constructor-style} induction. This leads to different handling of accessors, constructors, and the induction hypothesis (which in this case is a more general term).

\subsubsection{Generalizing Common Subterms}
\label{sec:GenCommon}
The algorithm for the generalization of common subterms proposed by Aderhold is quite similar to the Boyer-Moore generalization of minimal common subterms but with important differences. It is split into three steps: identifying generalizable subterms, generating proposals, and evaluating them.

The only difference when identifying generalizable subterms is that, in addition to the criteria in the Boyer-Moore generalization (i.e. be neither a variable, nor an explicit value template, nor an application of accessor functions), generalizable terms should not contain constructors. For instance, let us consider $(m\ \times\ n\ +\ n)\ +\ SUC(m)\ =\ (m\ \times\ n\ +\ m)\ +\ SUC(n)$. This clause occurs during the proof of the commutativity property of multiplication. The generalizable subterms are: $m\ \times\ n$, $m\ \times\ n\ +\ n$ and $m\ \times\ n\ +\ m$. Notice that the newly extended generalization criteria discard $(m\ \times\ n\ +\ n)\ +\ SUC(m)$ and $(m\ \times\ n\ +\ m)\ +\ SUC(n)$ as potentially generalizable subterms because they contain the constructor $SUC$, whereas in the Boyer-Moore generalization they would be accepted.

The second step generates proposals, which are sets of generalizable subterms that occur in a recursive position of a function or form one of the sides of an equation, eg.\ the subterm $a + b$ in equation $a + b = (a + b) + 0$. Proposals are filtered and only the ``suitable'' ones are kept by following an idea similar to the one in the Boyer-Moore system but with a different algorithm. A proposal is suitable for a formula $\varphi$ if the proposed terms are generalizable subterms of $\varphi$ and each occur at least twice in $\varphi$. Aderhold, also mentions a special check for equations, where the proposed term must also occur on both sides of the equation (the \emph{equation criterion}), or at least twice on one side. Having established that, each subterm of the formula is examined recursively for suitable proposals. Thus, in our first example, there is only a singleton, suitable proposal containing $m\ \times\ n$, which is proposed twice.

Two further differences in Aderhold's algorithm compared to the Boyer-Moore heuristic can be found in its third step. The Boyer-Moore system picks all of the minimal common subterms to generalize simultaneously (see Section \ref{sec:Gen}). Aderhold's algorithm only applies the single best proposal, after ordering these with respect to a number of criteria. The first criterion is the induction test: the induction scheme is used to test if an induction is possible on the generalized variable. A successful induction test shows that the proposal is much more likely to be correct. Other criteria include how often the proposal was made in the generation step and how many occurrences of the terms of the proposal can be found in the formula. After sorting the proposals, the first one is picked and applied. In VeriFun, the disprover is also used at this point to filter out over-generalizations. In our example, the generalized lemma produced by the single proposal $m\ \times\ n$ is $(n'\ +\ n)\ +\ SUC(m)\ =\ (n' +\ m)\ +\ SUC(n)$. It passes the induction test because an induction is possible on $n'$.

It is worth noting that, without some special machinery, the recursive nature of the waterfall model would defeat the purpose of only applying the best proposal. This is because the successfully generalized clause will be poured on top of the waterfall again and go through the same generalization heuristic which will essentially generalize the second proposal. Eventually all proposals will be generalized and not just the best one. We took special care to prevent this behaviour by using a technique similar to the warehouse filter (see Section \ref{sec:Loops}), storing the generalized terms and preventing a second generalization.

\subsubsection{Generalizing Variables Apart}
\label{sec:GenApart}
As indicated by one of Aderhold's Verifun examples \cite{aderhold}, it is often necessary to generalize apart the occurrences of $x$ in an expression such as $x+(x+x)=(x+x)+x$. In his algorithm, it is deemed necessary to rename the occurrences of the variable in the \emph{recursive} position of the functions involved. First, a heuristic filter is applied to detect the need for generalizing apart. The filter searches for a function $f$ and a variable $v$ that match the following criteria: $f$ should appear twice in the clause and $v$ should be an argument in the recursive position in the first appearance and an argument in a non-recursive position in the second appearance. If such a function and variable are found, the generalization of that variable is proposed. Two functions are used to ensure the variable is generalized in the correct positions of the clause. The variable is replaced in those positions by a fresh variable $v'$. A term $t$ is said to have been \textit{generalized apart successfully} if the whole term $t$ is replaced by $v'$ (i.e. $t=v'$) or at least one but not all occurrences of $v$ in $t$ were replaced by $v'$. For equations, it is required that both sides are generalized apart successfully. Once the generalization is applied, a check is used to verify if this is a \emph{useful} generalization. A useful generalization is one which was generalized successfully and in which all the equations were generalized apart successfully as well. A disprover is also used to rule out over-generalizations. Following this algorithm, our example is generalized to $n+(x+x)=(x+x)+n$.

If the first generalization proposal is not a useful generalization, another attempt is made. For all functions $g$, other than $f$, that appear in the clause and have the same recursive argument position as $f$, the variable is generalized apart in all such positions. This generalization is also checked for usefulness. This part of the algorithm accomplishes the generalization of an expression such as
	\[LENGTH(APPEND\ x\ x)\ =\ LENGTH\ x\ +\ LENGTH\ x
\]
 to 
	\[LENGTH(APPEND\ x'\ x)\ =\ LENGTH\ x'\ +\ LENGTH\ x
\]

At this point we should emphasize an important aspect of Aderhold's original algorithm. In his case, the algorithm allows for multiple recursive argument positions in functions. In fact a recursive position powerset is defined for each function, allowing for multiple definitions of the function with a different set of recursive argument positions for each definition. In our system, only functions with one recursive argument are allowed and hence only one position needs to be stored for each function. This simplifies the algorithm but the capability of the system to deal with different function definitions remains limited.

\subsubsection{Dealing with over-generalizations}
\label{sec:Counter}
Careful observation of several proof traces where the new algorithm did not contribute to a successful proof, combined with the fact that Aderhold uses a disprover to filter-out over-generalizations in several stages of the algorithms, lead to the implementation of a simple counterexample checker. 

For each generalized clause, a random example is generated for every free variable in it. Then simplification is used in an attempt to evaluate the grounded clause. The definitions of functions, constructors and accessors as well as some particular rewrite rules (such as $SUC\ 0\ =\ 1$ in order to deal with the HOL Light numeric $1$ that appears in some definitions) are given to the simplifier in order to accomplish this task. Additionally, we use a HOL Light conversion NUM\_REDUCE\_CONV to evaluate numeric expressions faster. This allows increased efficiency when handling terms that would otherwise take long to evaluate (such as terms containing exponential expressions). 

If the simplifier reduces the term to False, it disproves the clause and the generalization is rejected. Otherwise, if the term is reduced to True, the generalization is allowed to proceed. It is also worth noting that, in some cases there may not be enough rewrite rules to fully reduce the grounded term to either True or False. We have chosen the safe option, ie.\ to consider the corresponding clauses unsafe for generalization, and thus reject them.

As a simple illustration of our disprover in action, consider the clause $m\ +\ n\ =\ n\ +\ m$. The generalization apart algorithm attempts to generalize this to $m\ +\ n\ =\ n'\ +\ m$. The counterexample checker, however, can produce a counter example by instantiating $m$, $n$, and $n'$ to $SUC\ 0$, $0$, and $SUC(SUC(SUC\ 0))$ respectively. Then our simplifier is able to reduce the grounded clause $SUC\ 0\ +\ 0\ =\ SUC(SUC(SUC\ 0))\ + SUC\ 0$ to False and thus we can reject the overgeneralization.

In order to generate the random examples, we use the constructors defined in the Shell for the type. A ``maximum depth'' parameter is used to limit the size of the example. The constructors are applied randomly with a gradually increasing probability of using a bottom object. The same procedure is called for each constructor parameter. In the simple example of natural numbers, we have the option of using either $SUC$ or the bottom object $0$. In this case, $0$ has an increasing probability of being used and thus terminating the procedure.

Given that the counterexamples are generated randomly, often one random instance is insufficient to disprove a clause. In particular, for formulae that are falsified by few variable instantiations (such as $m\ \times\ n\ <\ m\ \times\ SUC\ n$ that is only false if $m\ =\ 0$) the counterexample checker will most likely fail to disprove them. Therefore, we apply multiple counterexample checks so as to achieve a more thorough (yet still incomplete) check. The number of such checks can be set by the user while taking into consideration the tradeoff between efficiency and thoroughness.
 
Usage of this counterexample checker altered the evaluation results significantly. The number of disproved clauses in every proof was added as a measure in our evaluation. The details of these results, along with all the others, are discussed in the next section.

\section{Evaluation}
\label{sec:Eval}
Our primary aim for the proper evaluation of such a system is to investigate its theorem proving potential as an automated tool within HOL Light. We also aim to evaluate the effect of our additions, including the loop elimination methods and new generalization techniques, on the performance of the system. There are considerably many parameters to take into consideration and various measures so an exhaustive evaluation of all scenarios is not possible. We describe the setup of our evaluation in Section \ref{sec:Setup} and we discuss the results in Section \ref{sec:Results}.

\subsection{Setup}
\label{sec:Setup}
Our evaluation involves inputting known theorems from existing theories into the system as conjectures and having it attempt to prove them fully automatically. In particular, we chose a total of 145 theorems from two test sets (see Appendix \ref{AB}). The first 120 form the basis of Peano arithmetic in HOL Light. The rest of the theorems were picked among the 50 examples from both Peano arithmetic and the list theory used for an evaluation of Rippling \cite{rippling}. The same test set is used by Aderhold for the evaluation of his generalization algorithms in VeriFun \cite{walther2002vfb}. It is worth noting that we were unable to test the whole set of 50 theorems, as some of them used functions that are not primitive recursive and thus cannot be defined within our system. The definitions of the functions used in our test sets that were added to the system are shown in Appendix \ref{AA}.

Deciding which parameters to test is important for the proper evaluation of the system. We first decided to consider six instances of the system. The first instance named ``BOYER MOORE'' (BM) is the pure reconstruction of Boulton's implementation with the addition of the counterexample checker. The second instance named ``BOYER MOORE EXT'' (BME) is the extension of the original implementation with all the additions we described in Section \ref{sec:Port} except from HOL Light's simplifier (see Section \ref{sec:HLsimp}) and the improved Generalization heuristic of Section \ref{sec:Aderhold}. We replaced the Boyer-Moore rewrite engine with the HOL Light simplifier (see Section \ref{sec:HLsimp}) to form the third instance of the system named ``BOYER MOORE REWRITE'' (BMR). The improved Generalization heuristic is tested in the fourth instance named ``BOYER MOORE GEN'' (BMG) where we substitute it for the generalization method in ``BOYER MOORE REWRITE''. After some result analysis, we tested a fifth instance of the system denoted by (BMG') which is the same as ``BOYER MOORE GEN'' except that it lacks the equation criterion (see Section \ref{sec:GenCommon} for a description of the criterion and Section \ref{sec:GeneralizationResults} for the reasoning behind its removal). Finally, having completed a detailed evaluation of our test sets, we combined the elements that were giving the best results into a final instance of the system called ``BOYER MOORE FINAL'' (BMF). Table \ref{tab:sysinst} shows the elements used in each of the six instances.
\begin{table}
	\centering
		\begin{tabular}{rcccccc}
			\ & BM & BME & BMR & BMG & BMG' & BMF \\ \hline \hline 
Basic Heuristics & x & x & x & x & x & x \\ \hline 
Counterexample checker & x & x & x & x & x & x \\ \hline 
Boyer-Moore simplifier & x & x & & & & \\ \hline 
HOL Light simplifier & & & x & x & x & x \\ \hline 
Warehouse filter & & x & x & x & x & x \\ \hline 
Maximum depth heuristic & & x & x & x & x & x \\ \hline 
Tautology heuristic & & x & x & x & x & x \\ \hline 
Setify heuristic & & x & x & x & x & x \\ \hline 
Boyer-Moore generalization & x & x & x & & & x \\ \hline 
Aderhold's generalization & & & & x & x & \\ \hline 
Variables apart generalization & & & & x & x & x \\ \hline 
Equation criterion & & & & x & & \\ \hline 
		\end{tabular}
	\caption{The six evaluated system instances}
	\label{tab:sysinst}
\end{table}

Another crucial aspect involved finding an appropriate setup for the numerous parameters that affect the system performance, given the fully automatic evaluation process. We decided to test the system with a minimum number of rewrite rules (see Appendix \ref{AA}) so as to have the least possible user intervention. The rules that were added are mainly properties of the involved datatypes and are not provable in the Boyer-Moore system (since they only involve constructors and accessors, not functions). Most of these properties are included in the datatype's shell. We also included a theorem involving the abbreviation of $SUC\ 0$ as $1$. Having the system automatically add rewrite rules depending on their potential usefulness in future proofs is an outstanding issue. Moreover, after some experimentation we decided that 5 counterexample checks per generalization attempt were sufficient to provide some useful results without a major impact on efficiency. Finally, after some observation of successful proofs, we chose a value of 12 for the maximum depth heuristic (see Section \ref{sec:Loops}). Clauses involved in successful proofs were never nearly as complex as those cut off by a maximum depth of 12.

There are also various measures that one could record to extract useful conclusions. We chose to log the result and the following four measures: 
\begin{enumerate}
	\item The time it takes for the system to prove a theorem (or to fail) as this is quite essential for this kind of systems.
	\item The number of proof steps as measured by the number of calls to a waterfall plus the number of inductions. The resulting number is proportional to the number of intermediate clauses produced and, upon success, proved by the system.
	\item The number of intermediate lemmas produced by generalization. Generalization is an unsafe operation, therefore having fewer generalizations is better for the system.
	\item The number of over-generalizations detected by the counterexample checker. This measure is used for the evaluation of the generalization techniques. Fewer over-generalizations indicate a better heuristic method.
\end{enumerate}

In addition to the above, we also examined the output of the generalization heuristic in relative detail as part of the evaluation. The clauses produced by generalization are separate lemmas speculated by the system (as opposed to the resulting clauses of the other heuristics which are simplifications or rewrites of the initial clause). Since these speculated lemmas often express interesting properties or theorems, they are investigated and evaluated separately. 

Our evaluation setup was implemented with the help of a wrapper function that recorded and gave the various measures as output. The data was collected in a spreadsheet and examined. At that point, we picked the most interesting or unexpected cases and examined them more closely in an attempt to analyse and explain them. Given the size of the test set and the numerous parameters that can be taken into consideration, one can extract a multitude of useful conclusions and ideas for the improvement of the system. Some of these are described in the following section.

\subsection{Results}
\label{sec:Results}

We begin with an evaluation of the Boyer-Moore system by discussing some general results in Section \ref{sec:GeneralResults}. This is followed  in Section \ref{sec:GeneralizationResults} by an analysis of the results obtained by having the improved generalization techniques in BMG when compared to BMR. We conclude our evaluation with a brief description of the results from BMF in Section \ref{sec:BMF}.

\subsubsection{General Results}
\label{sec:GeneralResults}
The first results from the tests showed that our reconstruction of Boulton's code worked as intended. We compared the results of BM with those originally given by Boulton \cite{boulton1992bma} and they matched. Moreover, based on our results we believe that the system can be a useful automated tactic for inductive proofs in HOL Light. The system was able to prove around 43\% of the 120 theorems in the first set and 33\% of the 25 theorems in the second test set automatically, ie.\ without any user interaction. An excerpt from the evaluation results containing successful proofs of BM is shown in Table \ref{tab:BMeval}. As another metric, if we look at a number of current HOL Light proofs (see Figure \ref{fig:proof}), we see that the same theorems are now proven automatically by BM or BMG in half a second or less. Therefore, we believe that it may prove useful as an automatic tactic in the hands of HOL Light users.

\begin{figure}
\ttfamily 
let LE\_SUC\_LT = prove \\
 (`!m n. (SUC m $\leq$ n) $\Leftrightarrow$ (m < n)`, \\
  GEN\_TAC THEN INDUCT\_TAC THEN ASM\_REWRITE\_TAC[LE; LT; NOT\_SUC; SUC\_INJ]);; \\
 \newline

let LT\_SUC\_LE = prove \\
 (`!m n. (m < SUC n) $\Leftrightarrow$ (m $\leq$ n)`, \\
  GEN\_TAC THEN INDUCT\_TAC THEN ONCE\_REWRITE\_TAC[LT; LE] THEN \\
  ASM\_REWRITE\_TAC[] THEN REWRITE\_TAC[LT]);; \\
 \newline

let LE\_LT = prove \\
 (`!m n. (m $\leq$ n) $\Leftrightarrow$ (m < n) $\vee$ (m = n)`,
  REPEAT INDUCT\_TAC THEN \\
  ASM\_REWRITE\_TAC[LE\_SUC; LT\_SUC; SUC\_INJ; LE\_0; LT\_0] THEN \\
  REWRITE\_TAC[LE; LT]);; \\
 \newline

let LT\_CASES = prove \\
 (`!m n. (m < n) $\vee$ (n < m) $\vee$ (m = n)`, \\
  REPEAT INDUCT\_TAC THEN ASM\_REWRITE\_TAC[LT\_SUC; SUC\_INJ] THEN \\
  REWRITE\_TAC[LT; NOT\_SUC; GSYM NOT\_SUC] THEN \\
  W(W (curry SPEC\_TAC) o hd o frees o snd) THEN \\
  INDUCT\_TAC THEN REWRITE\_TAC[LT\_0]);;
\caption{Some interactive HOL Light proofs that can be automated using either the ``BM'' or ``BMG'' tactics.}
\label{fig:proof}      
\end{figure}

\begin{table}
	\centering
		\begin{tabular}{lccccc}
		Theorem & Set & Time (s) & Steps & Inds & Gens \\
	\hline \hline
		\multicolumn{6}{c}{Proven by BM} \\
	\hline	m + n = n + m  & H & 0.047 & 19 & 3 & 0 \\
m + n + p = (m + n) + p  & H & 0.018 & 6 & 1 & 0 \\
m + n = m + p $\Leftrightarrow$ n = p  & H & 0.035 & 13 & 1 & 0 \\
m * n = n * m  & H & 0.194 & 48 & 7 & 1 \\
m * (n + p) = m * n + m * p  & H & 0.189 & 28 & 4 & 2 \\
SUC m $\leq$ n $\Leftrightarrow$ m $<$ n  & H & 0.065\ & 23 & 2 & 0 \\
m $<$ SUC n $\Leftrightarrow$ m $\leq$ n  & H & 0.179 & 54 & 4 & 0 \\
m $\leq$ n $\Leftrightarrow$ m $<$ n $\vee$ m = n  & H & 0.180 & 56 & 4 & 0 \\
(m + n) - (m + p) = n - p  & H & 0.098 & 16 & 2 & 1 \\
0 $<$ x EXP n $\Leftrightarrow$ $\neg$(x = 0) $\vee$ n = 0  & H & 0.337 & 58 & 4 & 2 \\
\hline \multicolumn{6}{c}{Proven by BMG}  \\
\hline LENGTH (REVERSE x) = LENGTH x & R &  0.057  & 15 & 2 & 1 \\
LENGTH (REVERSE (APPEND x y)) =  & R &  0.161  & 31 & 4 & 3 \\
LENGTH x + LENGTH y & & & & & \\
REVERSE (REVERSE x) = x & R &  0.071  & 17 & 2 & 1 \\
REVERSE (APPEND (REVERSE x) & R &  0.193  & 42 & 5 & 2 \\
 (REVERSE y)) = APPEND y x & & & & & \\
m $<$ n $\vee$ n $<$ m $\vee$ m = n  & H & 0.532 & 42 & 4 & 1 \\

		\end{tabular}
	\caption{The evaluation results (time, proof steps, inductions, and generalizations) for some successful proofs of BM and BMG. Set ``H'' corresponds to the HOL Light test set, whereas set ``R'' to the Rippling test set.}
	\label{tab:BMeval}
\end{table}

\paragraph{Looping examples} One of the most noticeable problems with the Boyer-Moore system in our initial evaluation runs was the sheer number of non-terminating examples. The BM instance of the system looped for more than a third of the cases that were tried. Some examples of theorems whose proofs loop in BM are shown in Table \ref{tab:loopeval}. This lead to the implementation of the warehouse filter and the maximum depth heuristic (see Section \ref{sec:Loops}) and the creation of the BME version of the system. The two procedures effectively reduced the number of looping examples in our two sets to 1\%. In particular, the maximum depth heuristic prevented 4 times as many loops as the warehouse filter. Although we are aware that the maximum depth heuristic may block proofs that would eventually succeed, we were unable to find such examples within our test sets.

\begin{table}
	\centering
		\begin{tabular}{l}
$\neg$EVEN n $\Leftrightarrow$ ODD n \\
EVEN n $\vee$ ODD n  \\
$\neg$(EVEN n $\wedge$ ODD n)  \\
EVEN (m + n) $\Leftrightarrow$ EVEN m $\Leftrightarrow$ EVEN n  \\
EVEN (m * n) $\Leftrightarrow$ EVEN m $\vee$ EVEN n  \\
EVEN (m EXP n) $\Leftrightarrow$ EVEN m $\wedge$ $\neg$(n = 0)  \\
ODD (m + n) $\Leftrightarrow$ $\neg$(ODD m $\Leftrightarrow$ ODD n)  \\
		\end{tabular}
	\caption{Examples of theorems that cause BM to loop, but can be proven automatically once $\neg ODD\ n \Leftrightarrow EVEN\ n$ is added as a rewrite rule.}
		\label{tab:loopeval}
\end{table}

\paragraph{Failed proofs} After careful investigation of some of the failed proofs, it was clear that the performance of the system could be greatly enhanced by properly managing the rewrite rule set manually. We observed that often a group of theorems could not be stratighforwardly proven because some simple lemmas were missing. For example, a number of theorems involving $EVEN$ and $ODD$, shown in Table \ref{tab:loopeval}, are provable by the system if we can demonstrate the theorem $\neg ODD\ n \Leftrightarrow EVEN\ n$ separately (eg.\ interactively without the waterfall) and add it to the rewrite rule set. This strengthens our view of the Boyer-Moore system as an automated procedure within an interactive theorem prover, where the user can manage the rewrite rule set properly so as to achieve optimal results.

\paragraph{Efficiency} Having timed the evaluation, we observed that the average proof time for successful proofs was under half a second for all five system instances and both test sets. Failed proofs (including those blocked by the loop elimination methods) took an average of 5 seconds, with a maximum of 25 seconds for BMR and 1 minute 15 seconds for BMG. If we consider 30 seconds as an acceptable time for an average user to expect a result in an interactive setting, these times are tolerable and provide enough room for more optimized cutoff heuristics, especially given the fact that successful proofs take considerably little time to complete. We also noted that HOL Light enhancements in BME compared to the original BM led to an average of 7\% fewer proof steps for successful proofs. For example, the lemma $m < SUC n \Leftrightarrow m \leq n$ is proven in 45 steps in BME as opposed to 54 in BM.

\paragraph{Comparing rewrite engines} The comparison between the results of BME and BMR is essentially a comparison between the original rewrite algorithm by Boulton and the usage of the HOL Light simplifier. BMR proved the same number of conjectures as BME. However, some small differences were observed in the efficiency of the two instances of the system. For BMR, there was a 6\% drop in the average number of proof steps in successful proofs as well as small drops in the number of inductions and generalizations. Even though the differences were small, they are still noteworthy as they are expected to scale up in larger proofs. The drop in the average number of inductions for successful proofs is an indication that some of the proofs required less inductions which, in turn, is a considerable advantage. Overall, using the HOL Light simplifier did not decrease the proof power of the system for the given test sets, but did offer a minor boost in the efficiency of the system.

\paragraph{Lemma speculation} The last important point which is indicative of the power of the system is the set of generalized terms. We filtered the speculated lemmas in successful proofs from BM and BMG. Examining the list of lemmas we can discover conjectures expressing interesting properties of our theory that are automatically speculated and proved. For natural numbers, these properties include commutativity of addition ($x+y=y+x$), associativity of multiplication $m\ \times (n\times p) =\ (m\times n)\ \times \ p$ and distributivity of multiplication over addition ($n\ \times \ p\ +\ m\ \times \ p\ =\ (n\ +\ m)\ \times \ p$) amongst many others.  A few trivial lemmas are speculated especially in BM because of the lack of the tautology checker which solves them before getting to generalization in BMG. To sum up, we observed that the system is capable of speculating interesting lemmas that may prove useful additions to the theory. This leads us to propose a filtering process at the end of a successful proof, which could heuristically select the most ``interesting'' theorems that were created by generalizations and make them available in the theory for the user to use in other proofs.

\subsubsection{Evaluating Generalization techniques within the Boyer-Moore system}
\label{sec:GeneralizationResults} 

Having established the potential of the system as an automated proof procedure, we investigated its usefulness when augmented with state-of-the-art generalization techniques such as the ones described in Section \ref{sec:Aderhold}. Results showed that on average 36\% of the successful proofs required one or more generalizations, so the importance and power of the generalization heuristic seems quite apparent. Unfortunately, the initial results with the new heuristic were not as expected, especially for the first test set. The success rate of BMG initially dropped significantly compared to BMR (29\% of the set proven compared to BMR's 44\%). Careful observation and result analysis was required to investigate the reasons for this somewhat unexpected decrease.

\paragraph{Rejecting over-generalizations} One of the immediately apparent problems of the new generalization heuristic was over-generalization that often led to non-theorems. This was mainly caused by generalizing variables apart. Noticing how Aderhold specifically mentions the necessity of a disprover, we implemented a simple counterexample checker (as described in Section \ref{sec:Counter}). The number of disproven generalizations then demonstrated some interesting facts. Primarily, BME and BMR made no overgeneralizations in any of the successful proofs. BMG's performance, however, increased to 37\% (compared to 29\% without the counterexample checker) and the measure showed an average 0.7 overgeneralizations per successful proofs. This made it clear that the counterexample checker is essential for the new generalization heuristic to work properly, since it often overgeneralizes. Examination of particular examples showed that in the vast majority problems were caused by the generalization of variables apart. For example, $n\leq n$ and $n\ \leq\ n\ \times\ n$ were both generalized to the non-theorems $n\leq n'$ and $n\ \leq\ n'\ \times\ n$. The counterexample checker is able to prevent both these overgeneralizations.

\paragraph{The Equation criterion} We were able to discover two particular cases where generalizing common terms should have been applied but is filtered out by our new generalization heuristic. In one of the cases, for instance, the clause $( m\ \times\ n\ =\ 0\ )\ \Leftrightarrow\ (m=0)\ \vee \ (n=0)$ is transformed into $n'\ \times\ n\ =\ 0\ \vee\ \neg (n'\ \times\ n\ +\ n\ =\ 0)\ \vee\ (n=0)$ after a few proof steps using the waterfall heuristics. At that point the Boyer-Moore generalization heuristic generalizes $n'\times n$ to $n''$, giving $n''\ =\ 0\ \vee \neg(n''\ +\ n\ =\ 0)\ \vee \ (n=0)$ which is then easily proved. However, the new heuristic based on Aderhold's approach does not allow this generalization. This is because $n'\times n$ appears only once on the left-hand side of the equation. According to the criterion for equations (see Section \ref{sec:GenCommon}), this generalization is ruled out and the system is unable to prove the original clause. However, in the given example this is a rational and useful generalization so empirically there's no reason why it should be ruled out. Having observed this issue we decided to rerun the evaluation for BMG while ignoring the equation criterion. BMG' had the same results as BMG with the addition of the proofs for the two problematic cases. There were no cases in our test sets where the lack of the equation criterion blocked the proof or led to an overgeneralization.

\paragraph{Comparing the heuristics} Even though BMR and BMG had similar results, rather surprisingly BMG appeared to be slightly less capable than BMR. We examined the particular cases where the Boyer-Moore generalization versus Aderhold's generalization produced different results. There was a number of theorems that BMR was able to solve and BMG failed. Examining these examples in detail showed that Aderhold's algorithm for generalizing common subterms rejected some crucial generalizations. As an example, it was unable to generalize $PRE(SUC\ (m\ +\ n)\ -\ m)\ =\ (m\ +\ n)\ -\ m$ which the Boyer-Moore generalization heuristic generalizes to $PRE(SUC\ n'\ -\ m)\ =\ n'\ -\ m$ and is then able to prove. Such a behaviour occurs because the procedure that generates the proposals for generalization in Aderhold's approach does not investigate deeper into constructors or accessors recursively, but only does so for functions and equations. Notably, there were a few cases where the new heuristic overgeneralized but the counterexample checker was unable to detect it. Finally, a few examples mostly in the second test set, were proven by BMG but BMR failed. Further investigation showed this success can be attributed to the generalization of variables apart. For example, the statement $DBL\ x = x\ +\ x$ is rewritten, using the definition of $DBL$: $DBL\ (SUC\ x)\ =\ SUC\ (SUC\ (DBL\ x))$, to $SUC\ (n\ +\ n)\ =\ n\ +\ SUC\ n$. BMR attempts to prove the latter by continuously applying induction until stopped by the maximum depth heuristic. In contrast, BMG proves this by generalizing $n$ apart resulting in $SUC\ (n'\ +\ n)\ =\ n'\ +\ SUC\ n$, which is then proven with a single induction on $n'$. Some more examples of theorems that demonstrate the differences between BMR, BMG and BMG' are given in Table \ref{tab:compeval}.

\begin{table}
	\centering
		\begin{tabular}{lcccl}
		Theorem & BMR & BMG & BMG' & Comment \\
		\hline
m * (n + p) = m * n + m * p  & Proved & Failed & Failed & \\ 
m + n $\leq$ m + p $\Leftrightarrow$ n $\leq$ p  & Proved & Failed & Failed \\ 
m * n = 0 $\Leftrightarrow$ m = 0 $\vee$ n = 0  & Proved & Failed & Proved & Equation criterion \\ 
m EXP n = 0 $\Leftrightarrow$ m = 0 $\wedge$ $\neg$(n = 0)  & Proved & Failed & Proved & Equation criterion \\ 
$\neg$(m $<$ n $\wedge$ n $<$ m)  & Failed & Proved & Proved & Variables apart gen \\ 
m $<$ n $\vee$ n $<$ m $\vee$ m = n  & Failed & Proved & Proved & Variables apart gen \\ 
		\end{tabular}
	\caption{Examples of theorems that demonstrate the differences in the results for BMR, BMG and BMG'. The comments refer to the main reasons behind these differences.}
	\label{tab:compeval}
\end{table}

\subsubsection{Combining the best components in BMF}
\label{sec:BMF}

Having completed a detailed analysis of our results, we were able to identify the components that were leading to the most successful proofs with the least possible steps. Since the results of BMR were slightly improved compared to BM, all the added components, including the loop detection heuristics and the HOL Light tools, were kept in BMF. Moreover, we concluded that the best choice for a generalization heuristic in our system is a combination of the original Boyer-Moore generalization heuristic with Aderhold's generalization of variables apart.

The evaluation of BMF using our two test sets showed some improvement over the original version (BM). In particular, BMF managed to prove 47\% of the lemmas in the first set and 37\% of the second set (as opposed to 43\% and 33\% respectively for BM), and these are the best results we were able to achieve so far with the given evaluation setup. Notably, BMF used fewer proof steps for successful proofs on average than BM (11\% reduction) and only slightly more inductions and generalizations (1-3\% on average). The complete evaluation results for BMF can be found in Appendix \ref{AB}.
\section{Related Work}
\label{sec:Related}
There is a variety of notable methodologies and tools that are used to achieve inductive proofs in modern theorem provers. We only provide a brief overview of some of the most notable ones next.

Proof Planning \cite{bundy1989sr} is a technique to guide the proof search. The basic idea is to attempt to construct a plan for the proof and then follow it to guide the proof itself. A few proof planners have been implemented such as the Oyster/Clam system \cite{bundy1990ocs}, the Omega system \cite{siekmann:pdomega} and IsaPlanner \cite{dixon2006ppf}. The latter is a proof planner implemented for Isabelle \cite{paulson1994igt}, which in turn is a generic theorem prover in ML. 

One of the main heuristic tools in Proof Planning is Rippling \cite{rippling}. Rippling gives a direction to the rewriting process. The idea is based on the observation that throughout the process, on each side of the equality being proved, there is an unchanging part (skeleton) and a changing part (wave-front). In principle the proof is guided towards rewrite rules that move the wave-front upwards in the syntax tree of the term. Rippling has proved to be a powerful heuristic for inductive proofs. However, there is still a possibility that the proof will block, thus requiring a ``patch'' or ``critic''. Critics include lemma speculations and generalizations, some of which can be produced automatically in the modern Proof Planning systems.

ACL2 \cite{kaufmann2000car} is a functional programming language and mechanical theorem prover based on Common LISP. It can deal with first order logic theorems and is particularly applicable to proofs about recursively defined functions and inductively constructed objects. The proofs in ACL2 are almost entirely automated. It requires, however some user hints either within the functions definitions or as helpful lemmas. These hints help guide the automated proof, which upon failure again relies upon the user to offer a solution. ACL2 is also capable of some generalizations based on the theory developed by Boyer and Moore (see Section \ref{sec:BM}). However, the development has been shifted mostly to the simplification step, where various decision procedures have been incorporated. 

As J. S. Moore explained\footnote{Personal communication.}, their focus throughout the evolution of the Waterfall model in ACL2 was to provide an automated system that uses simplification and induction to attempt a proof. If the system is unable to complete the proof, it is preferred that it fails early and provide detailed feedback and hints to the user so that he can unlock the proof. Our approach slightly differs from that advocated by the creators and developers of ACL2. We attempted to create a fully automated tactic as a tool to facilitate the otherwise interactive proofs in HOL Light. With that in mind, the system was designed and extended in a way that makes the proof search as exhaustive as possible. Nevertheless, our system is flexible enough to be able to accommodate and benefit from the methodologies and tools used in ACL2 and this could also be part of future work.

Finally, we also remark that there are other modern systems such as INKA \cite{hummel1990gig} and RRL \cite{kapur2000tps} and methodologies such as Inductionless Induction \cite{comon2001ii} under constant development for the support of inductive proofs. We refer the interested reader to the associated literature for more information about these.

\section{Future Work}
\label{sec:Future}
The encouraging results produced in a relatively limited timespan, provide multiple pointers for future work. The evaluation of the system was a time consuming process which, however, produced interesting results. We believe there is a lot of room for further evaluation of the system. On the one hand, one could expand the evaluation set using more theorems from different domains, eg.\ recursively-defined trees. On the other hand, one could delve deeper into the particular examples where the system failed or produced unexpected results and draw even more conclusions and ideas for improvement of the system.

There is also scope for improvement of the loop elimination heuristics. We have already considered possible measures, such as the number of clauses that remain to be proven in the pool of the waterfall and the number of inductions applied. We have also considered an incremental depth approach similar to the one used in the MESON tactic.

As far as the generalization heuristic is concerned, immediate future work would involve replacing the irrelevance heuristic by its counterpart from Aderhold's approach, known as ``inverse weakening''. Further experimentation for the optimal combination of criteria for generalizing common subterms within our system is also among our future plans.

\section{Conclusion}
\label{sec:Concl}
In this paper, we discussed the reconstruction and extension of the Boyer-Moore waterfall model for automated inductive proofs in HOL Light. An extensive and detailed evaluation of our implementation led to a plethora of useful and interesting conclusions about the relevance of the approach. Of those, the most important was the conclusion that the model, despite being over 30 years old, can improve the support for automated inductive proofs within HOL Light's interactive setting. Proofs, such as those in Fig.\ref{fig:proof}, were fully automated with a single usage of the Boyer-Moore tactic. Even though we were only able to prove 47\% of the evaluation set, if we keep in mind the simplicity and fully automated setup for the evaluation, this result is promising. In an interactive setup, the user will be able to manipulate various system parameters (see Section \ref{sec:UI}) so as to achieve optimal results. Often adding a simple rewrite rule may allow the proof to unblock. Moreover, the user will be able to stop a looping procedure manually even if our loop elimination heuristics fail. This is a common step during interactive theorem proving and is often used with automated tactics such as HOL Light's model elimination procedure MESON.

The flexible and highly programmable environment of HOL Light allowed the Boyer-Moore system to be fully integrated. It can interact with HOL Light's data structures and theorems seamlessly in the background and without burdening the user with the details of the underlying tactics being used. Moreover, we were able to implement various extensions that were easily integrated within the Boyer-Moore system thanks to its ``black-box'' heuristic approach. Our extensions helped make the system more user-friendly by improving its verbosity when desired and minimizing non-termination. Additionally, we achieved an increase in its performance by exploiting the tools and techniques available in HOL Light. 

Perhaps the most interesting results came from our attempt to apply state-of-the-art generalization techniques in our framework. We demonstrated the pros and cons of both the original Boyer-Moore and Aderhold's approach within this particular system. We showed, for example, that the original Boyer-Moore generalization of common subterms performs better in this context, whereas Aderhold's algorithm is hindered by the equation criterion and the specialised handling of constructors and accessors. However, we also showed that Aderhold's algorithm for generalization of variables apart can be a valuable heuristic in certain situations. This leads us to conclude that a combination of both approaches is required to achieve optimal results within our version of the Boyer-Moore model.

It is worth noting that the approach we followed strongly resembles the way the original Boyer-Moore prover evolved into ACL2. The initially simple waterfall model was upgraded into a complex system by various additions and extensions. Moreover, in ACL2 the user can provide theorems as ``hints'' for the automated procedure, similar to the way our system may require user intervention, eg.\ in the rewrite rule set, to achieve a proof. Therefore, based on our albeit more modest experiments, the way ACL2 evolved from the original waterfall model seems justified and natural and indicates that there is scope for developing an even more sophisticated inductive theorem proving system within HOL Light. We believe that the adaptable environment and our highly customisable implementation of the Boyer-Moore waterfall model provide sufficient grounds for our system to evolve into a significant tool for inductive proofs in HOL Light.

\clearpage
\appendix
\section{Definitions and Rewrite rules}
\label{AA}

Function definitions and basic rewrite rules used for the automated evaluation of our system.

\begin{figure}[htb]
	\centering
		\includegraphics[scale=.7]{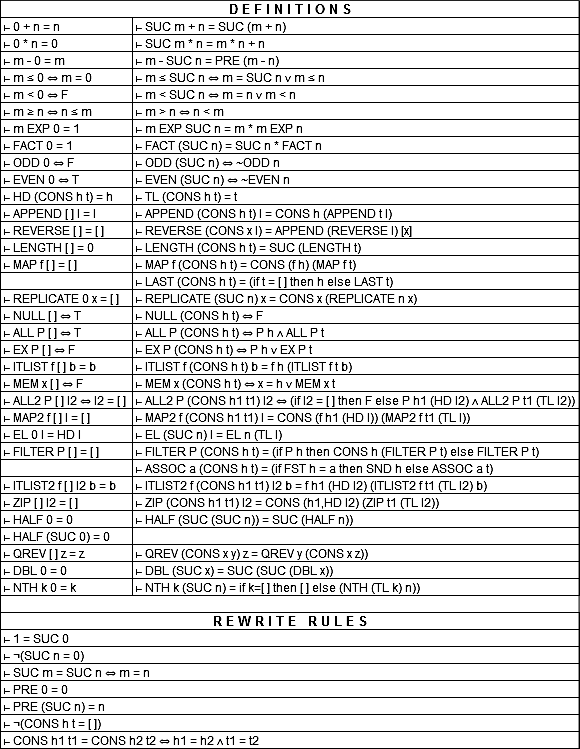}
	\label{fig:defs}
\end{figure}

\clearpage
\section{Evaluation results for BMF}
\label{AB}

Evaluations results for the final version (BMF) of our system.
These include whether the system was successful or not (\textit{false*} indicates failure by loop detection), the time in seconds (Time), the number of proof steps (Steps), inductions (Inds), generalizations (Gens), and detected overgeneralizations (Over).

\begin{figure}[ht]
	\centering
		\includegraphics[scale=.7]{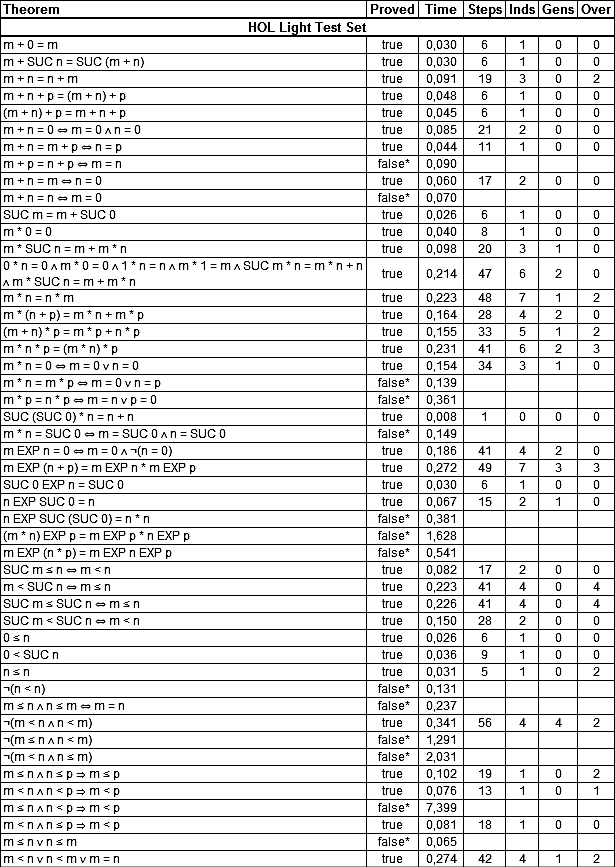}
	\label{fig:res1}
\end{figure}

\begin{figure}[ht]
	\centering
		\includegraphics[scale=.7]{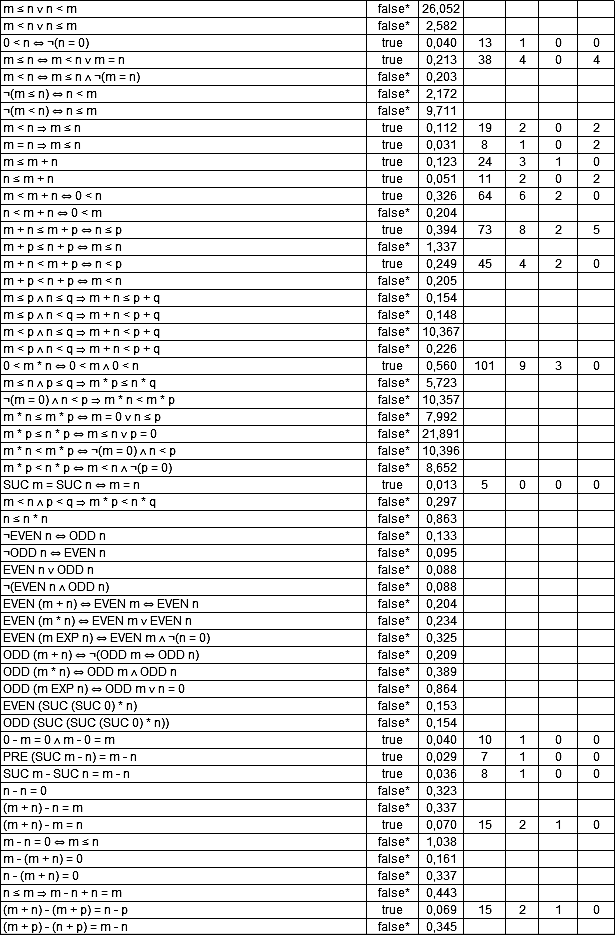}
	\label{fig:res2}
\end{figure}

\begin{figure}[ht]
	\centering
		\includegraphics[scale=.7]{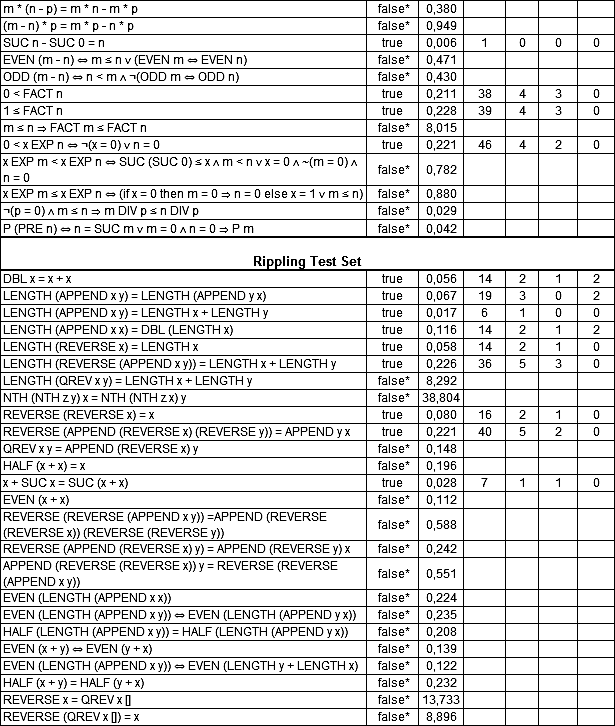}
	\label{fig:res3}
\end{figure}

\clearpage

% BibTeX users please use one of
%\bibliographystyle{spbasic}      % basic style, author-year citations
\bibliographystyle{spmpsci}      % mathematics and physical sciences
\bibliography{paper}   % name your BibTeX data base

\end{document}